\documentclass{aastex}
\usepackage{spr-astr-addons}
\usepackage{url}

\RequirePackage{color}

\begin{document}

\title{On the stability of planetary orbits in binary star systems \\ I. The S-type orbits}
\shorttitle{Stability of planetary orbits in binary stars}
\shortauthors{De Cesare et al.}

\author{G. De Cesare} 
\affil{INAF - Osservatorio di Astrofisica e Scienza dello Spazio, Via Piero Gobetti 93/3, I-40129 Bologna, Italy}
\and 
\author{R. Capuzzo-Dolcetta}
\affil{Department of Physics, Sapienza-Universit\`{a} di Roma, P.le A. Moro 5, I-00165 Rome, Italy}

\begin{abstract}
    Many exoplanets are discovered in binary star systems in internal or in circumbinary orbits. Whether the planet can be habitable or not depends on the possibility to maintain liquid water on its surface, and therefore on the luminosity of its host stars and on the dynamical properties of the planetary orbit. The trajectory of a planet in a double star system can be determined, approximating stars and planet with point masses, by solving numerically the equations of motion of the classical three-body system. In this study, we analyze a large data set of planetary orbits, made up with high precision long integration at varying: the mass of the planet, its distance from the primary star, the mass ratio for the two stars in the binary system, and the eccentricity of the star motion. To simulate the gravitational dynamics,  we use a 15th order integration scheme (IAS15, available within the REBOUND framework), that provides an optimal solution for long-term integration. In our data analysis, we evaluate if an orbit is stable or not and also provide the statistics of different types of instability: collisions with the primary or secondary star and planets ejected away from the binary star system. Concerning the stability, we find a significant number of orbits that are only marginally stable, according to the classification introduced by \cite{Musielak2005}. For planets of negligible mass, we estimate the \textit{critical}  semi-major axis $a_c$ as a function of the mass ratio and the eccentricity of the binary, in agreement with the results of \cite{howi99}. However we find that for very massive planets (Super-Jupiters) the critical semi-major axis decrease in some cases by a few percent, compared to cases in which the mass of the planet is negligible.
\end{abstract}


\section{Introduction}
\label{sec:intro}

   The first possible observational evidence of a Jupiter-like planet in the binary star system gamma Cephei was reported by \cite{Campbell1988}, based on the measurements of the radial velocity variations in a sample of stars. However, due to possibility that the low signal was due to chromospheric activities of the star, the existence of a planet in this system was subsequently questioned \citep{Walker1992}. It was thanks to more accurate measurements that the true nature of the gamma Cephei planet was finally confirmed \citep{Hatzes2003}, fifteen years after the first observation. It is worth noticing that the small distance (about $20$ AU) between the two stars in gamma Cephei has as a consequence a complex dynamics of the planetary orbit, whose stability is not guaranteed. Assuming the system as a classic Newtonian, isolated, three-body problem, no general analytical solution exists, also in the special case where the mass of the planet is negligible, i.e. for the so called {\em restricted three body problem}. A recent review of the three-body problem in the context of both historical and modern developments is presented by \cite{Musielak2014}. To date, the catalogue \footnote{\url{http://www.univie.ac.at/adg/schwarz/multiple.html}} maintained by \cite{Schwarz2016} reports $103$ binary and $26$ multiple confirmed star systems hosting planets, and $28$ candidates. The vast majority of multiple systems hosting planets are made up by three stars. In some case, the binary (or multiple) star system hosts more than one planet.

    Orbits of planets in binary systems are traditionally classified into three categories \citep{dvo86}: i) the Planet-type (P-type) external orbits around both stars in the binary, ii) the Satellite type (S-type) internal orbits, around one of the two stars, and iii) the Libration-type (L-type) orbits, corresponding to librations around the  Lagrangian equilibrium points L$_4$ and L$_5$, which are stable when the stellar mass ratio $m_1/(m_1+m_2)$ is less than $\sim 0.04$ (assuming $m_1 < m_2$). While L-type orbits are not normally of interest for exoplanets in binary systems, P-type and S-type orbits are of relevant astronomical impact and deserves to be deeply studied in their characteristics. For a recent review of planet formation and dynamical evolution in binary systems see \cite{Marzari2019}.

    The stability problem of planetary orbits in binary star systems have been investigated by many authors, with different methods and in different schemes. After pioneering work by \cite{hen68,hen69}, \cite{hengu70}, and \cite{sz80}, numerical approaches have been done by \cite{dvo86} (P-orbits in restricted assumptions),  \cite{1988A&A...191..385R} (S-orbits in restricted assumptions), and  \cite{pildvo02} (S-orbits with less restrictions). \cite{dvo86} and \cite{1988A&A...191..385R} numerically integrated a set of orbits by mean of a scheme based on Lie-series, as developed by \cite{de85} whose drawback is that of being limited to infinitesimal mass for the third-body, in what the primaries motion is assumed as the analytical solution of the two-body problem. The work done by \cite{pildvo02} using Bulirsh-Stoer algorithm, does not encounter in principle this limitation, but it was applied by the authors, again, to zero mass planets. Later on \cite{howi99}, hereafter referred to as H99, using a modified symplectic mapping technique initially developed by \cite{wiho91} to study the secular behavior of planetary motion around a dominating mass, investigated the planet orbital stability extending significantly with respect to previous works the time basis of integration; the authors assume that the mass of the planet is negligible. A straightforward Runge-Kutta-Fehlberg numerical technique was adopted by \cite{Musielak2005} who dealt with the stability of initially circular planetary orbits in both S- and P- configurations in the field of two stars revolving each other in circular motion. The Frequency Map Analysis (FMA) method \citep{Laskar1992} was for the first time applied by \cite{Turrini2004, Turrini2005} to investigate the dynamical stability of the giant planet in the $\gamma$ Cephei binary system. The FMA method was also applied by \cite{Marzari2016}, finding a significant number of stable planetary orbits in binaries beyond the critical distance from the primary star as evaluated by H99. \cite{Fatuzzo2006}, by performing a huge set of numerical integrations of Earth-like planetary orbits in binary systems, report the statistical distribution of the survival times of  planets in these systems. The authors conclude that in principle some unstable planets could have a sufficiently long survival time in the habitability zone. In systems with multiple planets orbiting around a single star, the minimum distance beyond which there are no close encounters is determined by the Hill stability limit \citep{Marcha1982, Gladman1993}. All these studies and results stimulated more specific works aimed on observational implications, including considerations on the habitability regions \cite[see, for instance][]{cu14,cu15}. 
    
    Just a few planetary systems are detected in star cluster, and so far only one (PSR B1620-26 b) in a globular cluster. The survivability and characteristics of planetary systems in star clusters was recently investigated by \cite{Cai2018, Cai2019, Elteren2019, Stock2020}. Among the observations of planetary systems in binaries, the case of $\nu$ Octantis is enigmatic since the planet is located outside the zone of orbital stability, as estimated by H99 for prograde orbits. To solve this puzzle, \cite{Eberle2010} suggested that the planet moves in a stable retrograde orbit. In a system (\emph{K2-290}) composed by three stars, retrograde motions with respect to the primary star's spin have been recently observed   \citep{Hjorth2021}.
    
     The secular motion has been considered in different works. In general, the classical analysis of secular motion is not accurate for objects in highly eccentric orbits; \cite{Michtchenko2004} introduce a semi-numerical approach to study the secular motion of two massive planets in co-planar orbits on high eccentricity (0.1 -0.6) orbits. \cite{Heppenheimer1978} introduced an analytical approximation for the three-body problem, providing a description of the secular orbital evolution for a planet in a binary star system. To assess the range in parameter space in which secular motion models provide accurate results,  \cite{Andrade-Ines2016} compared predictions for secular motion based on first and second order analytical models with $N$-body simulations. A semi-analytical correction to the \cite{Heppenheimer1978} solution was recently presented by \cite{Andrade-Ines2017}; this correction provides a quite accurate description of the S-type planetary secular motion in a given range of parameters.
    
    \cite{Quarles2020} recently performed a numerical study of the stability of S-type orbits following an approach similar to that of \cite{howi99} enlarging their initial parameter space. In particular, \cite{Quarles2020} probe the role of inclined planetary orbits by testing 4 inclinations for eccentric binaries, and 8 inclinations for circular binaries, integrating  the motion of Earth-mass particles with a symplectic scheme.  With respect to \cite{Quarles2020}, we probe a wider planet mass parameter space by simulating the motion of test particles and planets with 1 and 30 Jupiter-mass. Moreover, we provide a deeper characterization of the trajectories by recording, for unstable orbits, ejections or collisions with one of the two stars, and, for stable orbits by evaluating the average distance of the planet from the unperturbed orbit, i.e. the orbit that the planet would have without the secondary star.
    
    The paper is organized as follows. In Section \ref{sec:methods} we define the dataset made up with a huge number planetary orbits that we have numerically integrated at different initial conditions, and we describe the data analysis. In Section \ref{sec:results} we report on the results. We end up with the discussion in Section \ref{sec:discussion}.
    
\section{Method}
\label{sec:methods}	

\subsection{Code and dataset}
\label{subsec:numint}

    \begin{figure}[ht]
	\includegraphics[width=0.9\columnwidth]{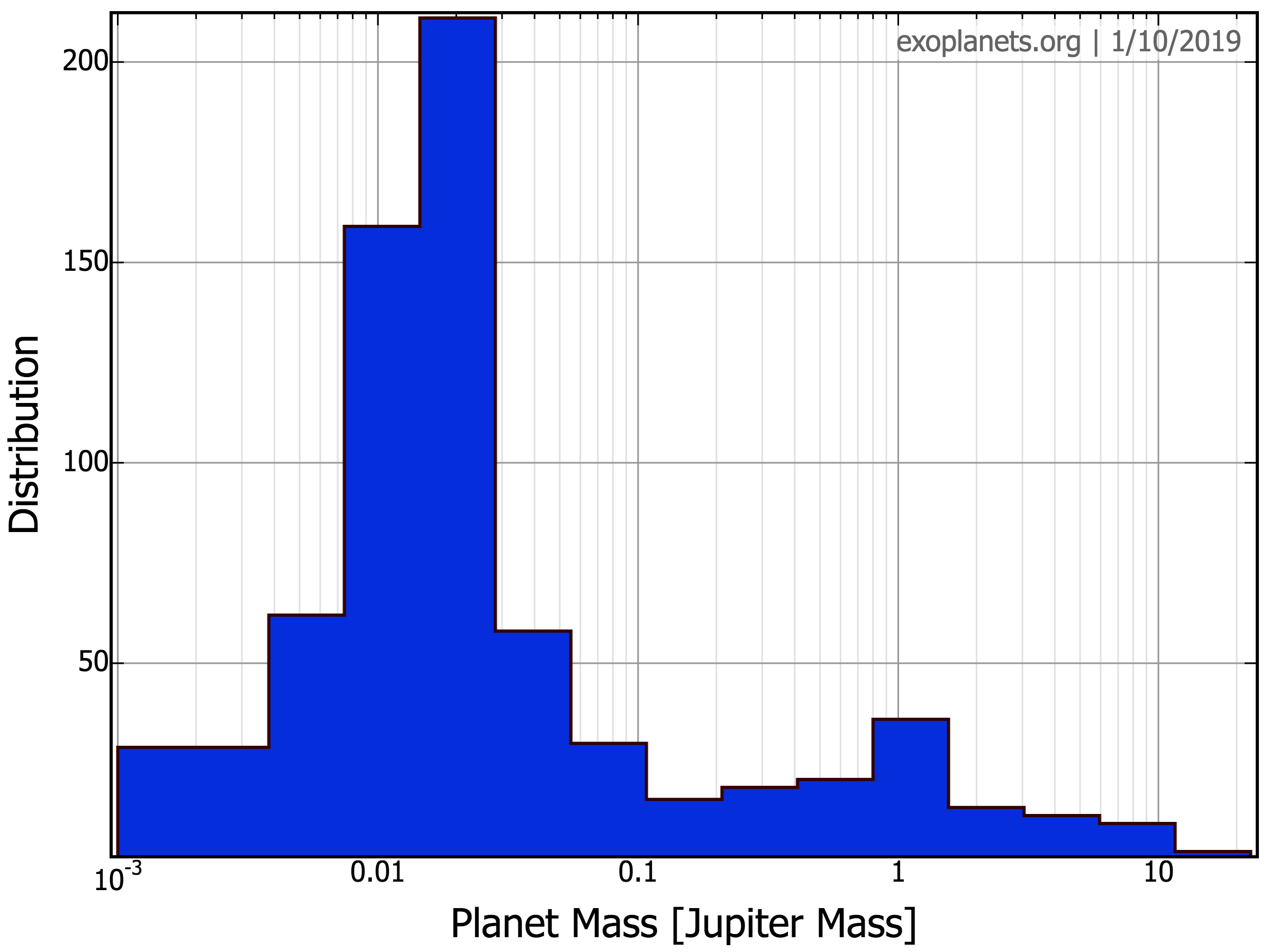}
    \caption{Distribution of the masses for the exoplanets discovered in binary star systems (from exoplanets.org). Although the majority of the planets have a small mass, a significant fraction of them has a mass equal to the mass of Jupiter or more.}
    \label{fig:masses}
\end{figure}
    
    We approach the problem of the planet orbital stability in a numerical way, by integrating the orbits under different initial conditions. We consider a system made up by two star $A$ (called primary) and $B$ (called secondary) in motion around their center of mass, with the planet in a co-planar orbit with zero initial eccentricity.  With \emph{unperturbed} orbit we will call the orbit around the primary star that the planet would have without the presence of the secondary star; in our configuration the  \emph{unperturbed} orbit is circular. Two examples of planetary orbits are shown in Fig. \ref{fig:orbit_cej} (top panel). We notice that the trajectory of the planet fills a ring nearby the \emph{unperturbed} orbit, whose "thickness" is a consequence of the gravitational field of the secondary star.

    Many ordinary differential equations (ODEs) integration methods can be applied to calculate the orbits of a gravitational N-body system \citep{Aarseth2003}. As test particles do not interact with each other, H99 simulate large numbers of single planet systems in one single integration. For finite mass planets this approach is not possible and each simulation must be managed individually. H99 use a mixed approach for the numerical integration of the orbits. When the planet is near the primary star, the symplectic integration scheme \cite{wiho91} is applied. This technique is accurate for motion around a single star, but also in the case of a binary star system it provides a reasonable approximation, the gravity of the primary star being dominant. Far from this condition, a Bulirsh-Stoer (\emph{BS}) integration method is applied;  this  algorithm combines a fairly accuracy at a relatively little computational cost \citep{NR}.  To achieve a very accurate integration of the orbits, in this study we will use the  high-precision integration scheme \emph{IAS15} \citep{Rein2015}  throughout all the duration of the motion.  With \emph{IAS15}, a 15th-order integrator, the systematics errors are kept below machine precision for long term integration over at least $10^9$ orbital time scale. \emph{IAS15} is an integration options available in REBOUND \cite{Rein2012}, a modern open source code for gravitational dynamics. REBOUND is written in standard C99; an easy to use and convenient python wrapper is also provided.

    The planetary system is made up by two stars: \emph{primary star} $A$, and the \emph{secondary star} $B$, in elliptical  motion in the center of mass frame of the two stars. The distance of the planet from the \emph{primary star} plays a crucial role on the stability. Generally speaking, we expect that the farther the planet is from the primary star, the stronger is the perturbation due to the secondary star, so that the orbit can eventually become unstable. The \emph{unperturbed} circular orbit is characterized by the initial (transverse) speed of the planet $v_P$  given by: 
    
    \begin{equation}
        v_P = \sqrt{G\frac{m_A + m_P}{r}},
    \label{eq:velocity}
    \end{equation}
    
    where $r$ is the initial distance of the planet from the primary star, $m_P$ is the mass of the planet and $m_A$ is the mass of the primary star.  Following H99, we consider as free parameters the \emph{mass ratio} and the eccentricity of the binary star system, and the initial semi-major axis of the planet orbit. The \emph{mass ratio} is defined as:
    
    \begin{equation}
        \mu = \frac{m_B}{m_A + m_B},
    \label{eq:mass_ration}
    \end{equation}
    
    For equal mass stars, $\mu = 0.5$. The smaller the $\mu$ the lower the gravitational perturbation due to the secondary star.  As in the study of H99, in our dataset the mass ratio of the binary $\mu$ goes from $0.1$ to $0.9$ (with $\Delta \mu = 0.1$) and the eccentricity $e$ from $0$ to $0.8$ (with $\Delta e = 0.1)$. The semi-major axis of the binary is equal to $1$ AU, and the binary initial phase is at periapse or at apoapse. Unlike the study of H99, where only planets with a negligible mass are considered, we also study the motion of planets with finite non negligible mass. We made numerical integration for a planet of infinitesimal mass and of mass equal to 1 and 30 Jupiter masses, which is a reasonable choice, taking into account the mass distribution of observed for planets in binary systems (Fig. \ref{fig:masses}). By putting all these degrees of freedom together, our dataset is made by more than ten thousand orbits. Concerning the masses of the planets, \cite{Stevens2013} proposed a general classification based on the following mass regimes: sub-Earths ($10^{-8} M_\Earth$ - $0.1 M_\Earth$), Earths/Super-Earths ($0.1 M_\Earth$ - $10 M_\Earth$), Neptunes ($10 M_\Earth$ - $100 M_\Earth$), Jupiters ($100 M_\Earth$ - $10^3 M_\Earth$), Super-Jupiters ($10^3 M_\Earth$ - $13 M_{Jup}$), brown dwarfs ($13 M_{Jup}$ - $0.07 M_\Sun$) and stellar companions ($0.07 M_\Sun$ - $1 M_\Sun$).
    
    The integration time extension is an important parameter of the numerical study. Actually,  long-term integration provides a greater confidence about the stability of a planetary orbit, but this obviously implies a cost in term of computational time. The duration of the integration should be a trade-off between the duration of the numerical integration and the accuracy of the results. Following the studies of \cite{howi99}, an integration time equal to $10\,000$ star orbits it is sufficient to obtain reliable results. 

\subsection{Analysis of the orbits}
\label{subsec:stability}  

    We classify the simulated orbits on the basis of their stability according to the following definitions. We consider an orbit as unstable if (i) the planet will \textit{collide} in a finite time with one of the two stars or (ii) it is \textit{ejected} from the binary system. We catch a collision if the star-planet distance reduces to a solar radius, an ejection if the planet moves away from the center of mass of the binary system by more than $100$ AU. In these cases we record the orbit lifetime (aka survival time), that is the time elapsed from the beginning of the orbit to the collision or the escape. In this classification, we do not explicitly refer to chaos, i.e. that sensitivity to initial conditions that makes impossible an accurate long term prediction of the position and velocity of the system particles. Orbits that we classify as stable, can have chaotic behavior. In any case, all the stable orbits we find in our sample are well confined in a ring surrounding the unperturbed orbit (Fig. \ref{fig:orbit_cej}). In general, stable orbits are characterized by how much they deviate from the unperturbed orbit; the orbits close to the primary star are almost circular, and little affected by the gravity of the secondary star. A simple way to characterize the stable orbits is through the difference $\Delta r$ of the distance $r$ of the planet from the primary star with respect to the initial value mediated along the orbit:
    
    \begin{equation}
        \langle \Delta r\rangle_T = \frac{\langle r(t) - r_i \rangle}{r_i},
    \label{eq:delta}
    \end{equation}
    
    where the average is made over the entire simulated orbit of the planet, so T corresponds to $10\,000$ orbital periods the binary. Hereafter this mean is referred simply as $\Delta r$.    
    Of course, $\Delta r = 0$ only if there is no interaction with the secondary star ($m_B = 0$). \cite{Musielak2005} classified the planetary orbits in binary on the basis of the value of the above parameter (Eq. \ref{eq:delta}):  (1) stable if $\Delta r \le 5 \%$,  (2) marginally stable if  $5 \% < \Delta r \le 35 \%$, (3) unstable if $\Delta r > 35 \%$.  The $5 \%$ threshold for stability is motivated by some studies (e.g. \cite{Kasting1993}, \cite{Underwood2003}) which show that this limit is required for the Earth to remain in the habitable zone and therefore allow the evolution of life.
    
\section{Results}
\label{sec:results}

    We report in this section how the stability depend on the mass of the planet, the initial semi-major axis of the planetary orbit, and on the mass ratio and eccentricity of the binary star system. 
    
\subsection{Collisions and ejections}
\label{sec:collisions_ejections}

    \begin{figure*}
	\begin{tabular}{cc}
	\includegraphics[width=0.96\columnwidth]{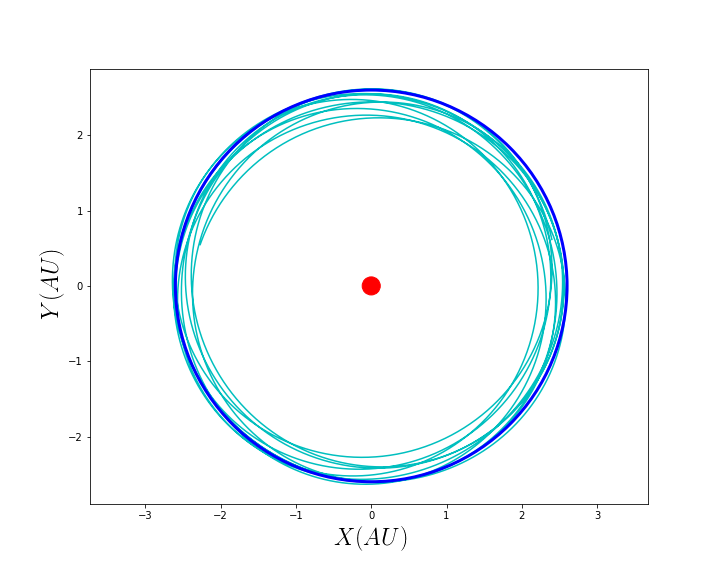}&
    \includegraphics[width=0.96\columnwidth]{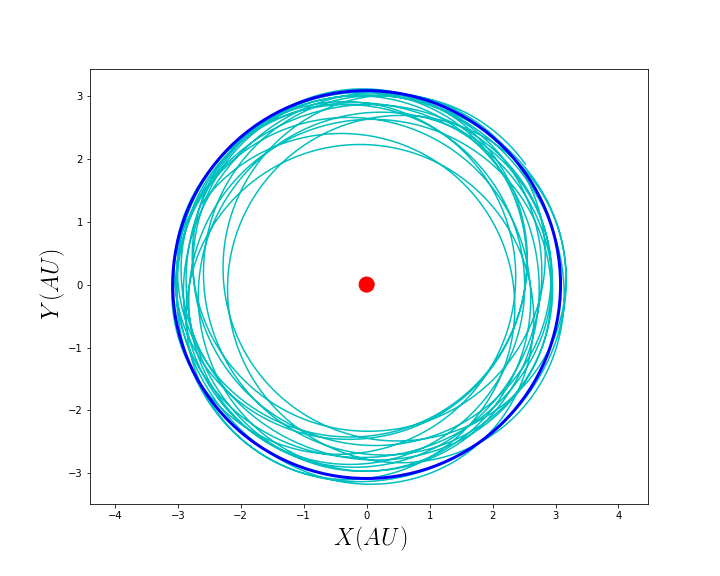}\\  
    \includegraphics[width=0.96\columnwidth]{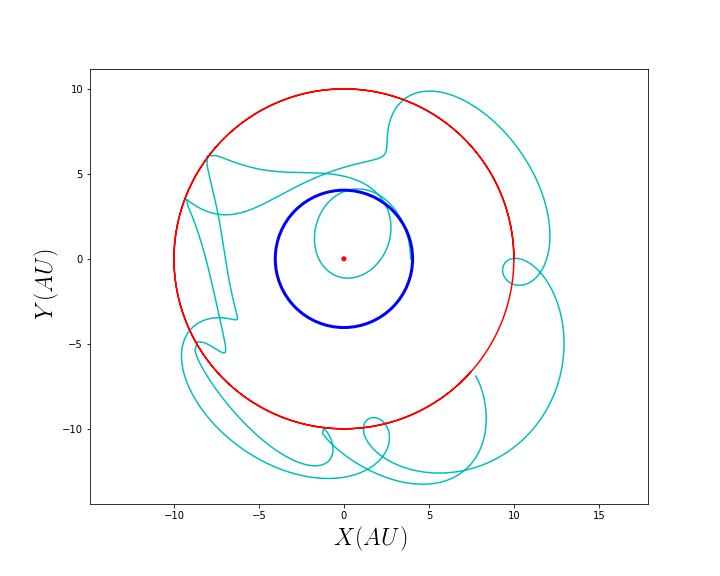}&
    \includegraphics[width=0.96\columnwidth]{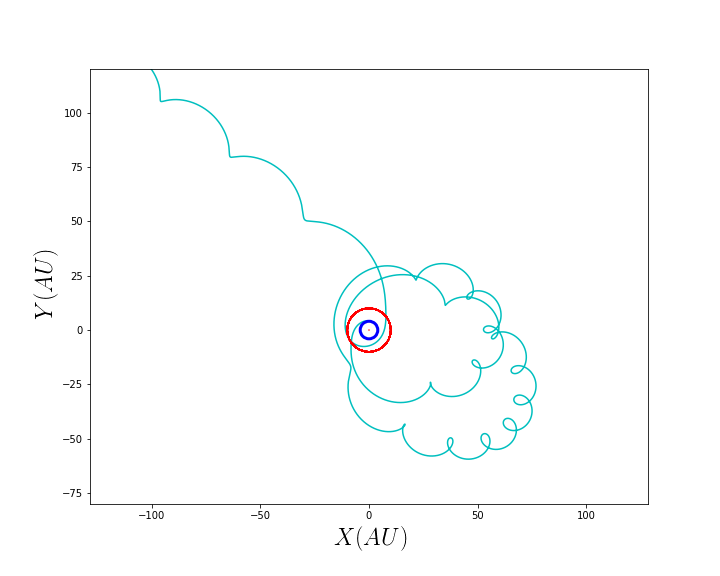}
    \end{tabular}
    \caption{Top left and right panels: a stable ($\Delta r = 3.9\% $), and a marginally stable ($\Delta r = 7.9\% $) orbit (see Section \ref{subsec:stability}). Bottom left, right panels: a collision of the planet with the secondary star, and an ejection of the planet away from the binary. The red circle is the orbit of the secondary star in the frame centered in the primary, the blue circle is the orbit of the unperturbed planetary motion.}
    \label{fig:orbit_cej}
\end{figure*}
   
    \begin{figure}
	\includegraphics[width=\columnwidth]{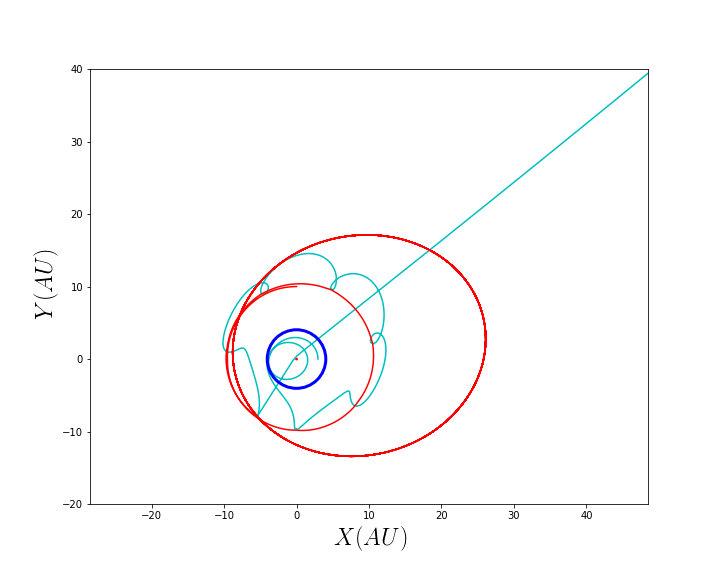}
    \caption{A peculiar unstable orbit of a planet of 30 Jupiter masses. At first, the planet moves away from the main star to be captured by the gravitational field of the secondary star and ejected away from the system. The orbit of the secondary star is significantly modified into an elliptic orbit.}
    \label{fig:peculiar-orbit}
\end{figure}
    \begin{figure}
	\includegraphics[width=\columnwidth]{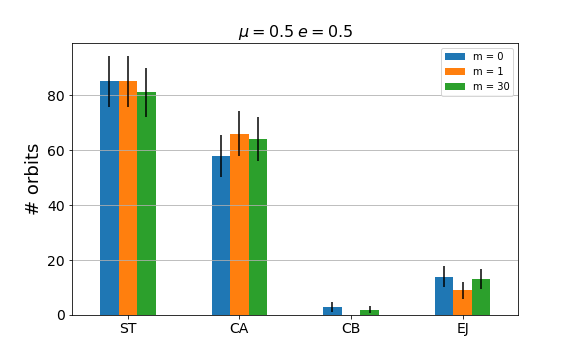}
    \caption{Synopsis of stability of the studied orbits for three different values of  the planet mass ($0$, $1$, $30$ Jupiter mass) in a binary with $\mu =0.5$, $e = 0.5$. ST indicate stable and marginally stable orbits, CA and CB indicate orbits leading to a collision with the primary or secondary star. Orbits leading the planet to be ejected away from the binary star system are labelled with EJ. Error bars represent $1\,\sigma$ confidence intervals.}
    \label{fig:stat}
\end{figure}
   
    We introduce with some examples\footnote{In the examples shown, the distance between the two stars is equal to $10$ AU} the statistics of possible collisions of the planet with one of the two stars of the binary and of ejections from the binary system. As a demo, we built a real-time animation based on a 4th order Runke-Kutta integrator\footnote{Available at the url  \url{https://giovixo.github.io/planets/}; the source code can be accessed with the DOI \url{https://zenodo.org/record/4014681}}. The two orbits shown on the top left and right panel of Fig. \ref{fig:orbit_cej} are classified as \emph{stable} and \emph{marginally stable}, on the basis of their estimated $\Delta r$ (see Section \ref{subsec:stability}). As the full planetary orbit is too long to be displayed effectively, the plot is obtained integrating the final part of motion within a time interval $T = 30\, 000\,$ days. We notice that despite $\Delta r$ of the two orbits differ significantly, the planetary orbit shows similar characteristics.

    The bottom left panel of Fig \ref{fig:orbit_cej} shows a collision with the secondary star. The orbit is the result of the numerical integration from the initial time to the star collision. The bottom right panel of Fig. \ref{fig:orbit_cej} shows a orbit where the planet is ejected away from the binary. It is interesting to notice that, before escaping, the planet moves in an outer orbit. This leads to the hypothesis that, under certain conditions, a transition from an internal (S-type) orbit to an external (P-type) orbit could occur. An interesting orbit is shown in Fig. \ref{fig:peculiar-orbit}. In this case, before being ejected away from the binary system, the massive planet scatters with the secondary star, producing a significant perturbation of the stellar orbit. As the consequence of the interaction planet-star, the star moves from a substantially circular orbit to a wider elliptical orbit.

    Figure \ref{fig:stat} reports the statistics of planet-star collision and ejection events for a binary with $e = 0.5$ and $\mu = 0.5$. In this particular case, the number of collisions and ejections do not depend on the planetary mass.
    
\subsection{Stability of the orbits}
\label{sec:stability}

    \begin{table*}
    \centering
    \begin{tabular}{|c| c c c c c c c c |}
    \multicolumn{9}{c}{Planet of negligible mass, $\mu = 0.5$, $e = 0.5$ } \\
    \hline 
    \hline
    \multicolumn{1}{|c}{} & \multicolumn{8}{c|}{Binary Initially at Periapse} \\
    \hline
    \bf{$a$ (AU)}  &  P1       & P2        & P3        & P4       & P5      & P6      & P7        & P8 \\
    \hline
    0.17      & CA(4)   &  CA($<1$) & CA($<1$)  &  CB(1)   & CA(1)   & CA(1)   & CA($<1$)  & EJ(1)         \\ 
    0.16      & EJ(4)   &  CA(2)    & CA($<1$)  &  EJ(1)   & CA(1)   & CA(2)   & EJ(6)     & CA($<1$)       \\
    0.15      & CA(4)   &  CA(5)    & CA($<1$)  &  CA(2)   & CA(6)   & CA(2)   & EJ(2)     & CA(2)           \\
    0.14      & CA(86)  &  CA(14)   & EJ(11)    &  CA(5)   & CA(135) & CA(16)  & CA(1)     & EJ(8)            \\
    0.13      & CA(221) &  CA(193)  & CA(11)    &  EJ(205) & 3.6     & CA(139) & CA(38)    & EJ(23)            \\
    \cline{2-9}
    \bf{0.12} & 7.7     &  7.3      & 8.1       &  5.8     & 3.5     & 5.9     & 7.7       & 8.0            \\
    0.11      & 6.4     &  6.1      & 5.7       &  4.2     & 3.8     & 4.2     & 5.7       & 6.1             \\
    0.10      & 5.7     &  5.4      & 5.2       &  4.2     & 3.9     & 4.2     & 5.2       & 5.4              \\
    0.09      & 5.2     &  4.9      & 4.8       &  4.2     & 4.0     & 4.2     & 4.8       & 4.9               \\
    0.08      & 4.7     &  4.5      & 4.4       &  4.1     & 3.9     & 4.1     & 4.4       & 4.5                \\
    \hline
    \end{tabular}
    \caption{Characterization of the planetary orbits for a given value of mass ratio $\mu = 0.5$, and eccentricity of the binary $e = 0.5$. 
    $a$ is the initial major semi-axis or the planetary orbit. The calculation is made for 8 equispaced longitudes (P1-P8) with the binary initially at periapse. Each cell in this table is associated with a complete simulation of a planetary orbit. The critical semi-major axis is $a_c = 0.12$. For stable orbits ($a \le 0.12$) the value of $\Delta r$ is reported. For unstable orbits, we use a code to indicate whether the instability is due to a collision with the primary or secondary star (CA, CB) or if the planet is ejected away (EJ). The survival time of the orbit in units of 10 binary periods is quoted in brackets. }
    \label{tab:stability_p}
\end{table*}

\begin{table*}
    \centering
    \begin{tabular}{|c| c c c c c c c c |}
    \multicolumn{9}{c}{Planet of negligible mass, $\mu = 0.5$, $e = 0.5$ } \\
    \hline 
    \hline
    \multicolumn{1}{|c}{} & \multicolumn{8}{c|}{Binary Initially at Apoapse} \\
    \hline
    \bf{a (AU)}    &  A1       & A2        & A3    & A4        & A5     & A6        & A7      & A8      \\
    \hline
    0.17      &   CA(3)   &  CA(1)   & EJ(3)  & CA($<1$) & CA(2)   &   CA(3)   & CA(2)   & CA($<1$)  \\
    0.16      &   CA(2)   &  CB(1)   & EJ(3)  & CA(1)    & EJ(6)   &   CA(3)   & CA(1)   & EJ(3)      \\
    0.15      &   CA(1)   &  CA(9)   & CA(2)  & CA(1)    & CA(9)   &   CA(3)   & CA(1)   & CA(1)       \\
    0.14      &   2.3     &  CA(36)  & EJ(50) & CA(15)   & CA(16)  &   CA(34)  & CA(10)  & CA(20)       \\
    0.13      &   CA(235) &  CA(665) & 5.5    &  6.6     & CA(681) &   CA(149) & CA(770) & 5.8           \\
    \cline{2-9}
    \bf{0.12} &   4.8     &  4.9     &  5.3   &  5.2    & 6.4      &   5.4     & 5.1      & 5.1 \\
    0.11      &   4.8     &  4.8     &  4.9   &  5.0    & 5.0      &   5.0     & 4.9      & 4.8    \\  
    0.10      &   5.0     &  5.0     &  4.9   &  4.9    & 4.8      &   4.9     & 4.9      & 5.0     \\  
    0.09      &   4.7     &  4.8     &  4.8   &  4.7    & 4.7      &   4.7     & 4.8      & 4.8      \\  
    0.08      &   4.5     &  4.5     &  4.5   &  4.5    & 4.5      &   4.5     & 4.5      & 4.5       \\  
    \hline
    \end{tabular}
    \caption{Characterization of the planetary orbits for a given value of mass ratio $\mu = 0.5$, and eccentricity of the binary $e = 0.5$ (see Table \ref{tab:stability_p}).  The calculation is made for 8 equispaced longitudes (A1-A8) with the binary initially at apoapse.}
   
    \label{tab:stability_a}
\end{table*}

    \begin{table*}
    \centering
    \begin{tabular}{|c c c c c c c c c c c|}
    \multicolumn{11}{c}{Critical semi-major axis ($m_p = 0, 1, 30$ M$_\textrm{J}$)} \\
    \hline 
    \multicolumn{6}{|c}{} & \multicolumn{1}{c}{$\mu$} & \multicolumn{4}{c|}{} \\
            &    & 0.10        & 0.20       & 0.30        & 0.40         & 0.50         & 0.60       & 0.70       & 0.80        & 0.90 \\
    $e$ & $m$  (M$_\textrm{J}$) & \multicolumn{8}{c}{} & \multicolumn{1}{c|}{} \\   
    \cline{3-11}      
    0.0     & 30 & \textbf{0.43}        & \textbf{0.36}  & \textbf{0.38}  & \textbf{0.31}  & \textbf{0.27} & 0.23  & 0.20  & \textbf{0.17} & 0.13 \\
            &  1 & 0.45                 & \textbf{0.37}  & 0.37           & 0.30           & 0.26          & 0.23  & 0.20  & 0.16          & 0.13    \\
            &  0 & 0.45                 & 0.38           & 0.37           & 0.30           & 0.26          & 0.23  & 0.20  & 0.16          & 0.13 \\
    \cline{3-11}
    0.1     & 30 & 0.37        & \textbf{0.31}  & \textbf{0.30} & \textbf{0.28} & 0.24         & 0.21       & 0.18       & 0.15        & 0.11 \\
            &  1 & 0.37        & \textbf{0.34}  & \textbf{0.30} & 0.27          & 0.24         & 0.21       & 0.18       & 0.15        & 0.11 \\
            &  0 & 0.37        & 0.32           & 0.29          & 0.27          & 0.24         & 0.21       & 0.18       & 0.15        & 0.11 \\
    \cline{3-11}
    0.2     & 30 & 0.32        & 0.27       & \textbf{0.26}        & 0.23         & \textbf{0.20}   & 0.19       & 0.16       & 0.13        & 0.10  \\
            &  1 & 0.32        & 0.27       & \textbf{0.26}        & 0.23         & \textbf{0.20}   & 0.19       & 0.16       & 0.13        & 0.10  \\
            &  0 & 0.32        & 0.27       & 0.25                 & 0.23         & 0.21            & 0.19       & 0.16       & 0.13        & 0.10  \\
    \cline{3-11}
    0.3     & 30 & \textbf{0.26} & 0.24   & 0.21        & 0.19         & 0.18          & 0.16       & 0.14       & 0.12        & 0.09 \\
            &  1 & 0.28          & 0.24   & 0.21        & 0.19         & \textbf{0.17} & 0.16       & 0.14       & 0.12        & 0.09   \\
            &  0 & 0.28          & 0.24   & 0.21        & 0.19         & 0.18          & 0.16       & 0.14       & 0.12        & 0.09 \\
    \cline{3-11}
    0.4     & 30 & \textbf{0.21} & 0.20       & 0.18        & 0.16         & 0.14         & 0.13       & \textbf{0.12} & 0.10       & 0.07 \\
            &  1 & 0.23          & 0.20       & 0.18        & 0.16         & 0.14         & 0.13       & 0.11          & 0.10       & 0.07  \\
            &  0 & 0.23          & 0.20       & 0.18        & 0.16         & 0.14         & 0.13       & 0.11          & 0.10       & 0.07 \\
    \cline{3-11}
    0.5     & 30 & \textbf{0.16}        & \textbf{0.15} & 0.14        & 0.13         & 0.12         & 0.10       & 0.09        & 0.08       & 0.06  \\
            &  1 & \textbf{0.19}        & 0.16          & 0.14        & 0.13         & 0.12         & 0.10       & 0.09        & 0.08       & 0.06    \\
            & 0  & 0.18        & 0.16          & 0.14        & 0.13         & 0.12         & 0.10       & 0.09        & 0.08       & 0.06  \\
    \cline{3-11}
    0.6     & 30 & \textbf{0.12}  & \textbf{0.11} & \textbf{0.10} & \textbf{0.09}  & 0.09  & 0.08       & 0.07        & 0.06       & \textbf{0.050}   \\
            &  1 & 0.13           & \textbf{0.11} & 0.11          & 0.10           & 0.09  & 0.08       & 0.07        & 0.06       & \textbf{0.050}    \\
            &  0 & 0.13           & 0.12          & 0.11          & 0.10           & 0.09  & 0.08       & 0.07        & 0.06       & 0.045 \\
    \cline{3-11}
    0.7     & 30 & \textbf{0.08} & 0.08       & 0.07        & \textbf{0.07}  & 0.06         & 0.05       & 0.05        & 0.045      & 0.035 \\
            &  1 & 0.09          & 0.08       & 0.07        & \textbf{0.07}  & 0.06         & \textbf{0.06}       & 0.05        & 0.045      & 0.035   \\
            &  0 & 0.09          & 0.08       & 0.07        & 0.06           & 0.06         & 0.05       & 0.05        & 0.045      & 0.035 \\
    \cline{3-11}
    0.8     & 30 & 0.05        & \textbf{0.04} & 0.04        & 0.04         & 0.04         & 0.035      & 0.03        & 0.025      & \textbf{0.0235}  \\
            &  1 & 0.05        & 0.05          & 0.04        & 0.04         & 0.04         & 0.035      & 0.03        & 0.025      & 0.0230  \\
            &  0 & 0.05        & 0.05          & 0.04        & 0.04         & 0.04         & 0.035      & 0.03        & 0.025      & 0.0230 \\
    \hline
    \end{tabular}
    \caption{Critical semi-major axis $a_c$  for a planet  mass ($m$) of 1, 30 Jupiter mass (M$_\textrm{J}$), and for a planet of negligible mass (test particle), as a function of the mass ratio $\mu$ and the eccentricity $e$ of the binary. The text is in boldface when the $a_c$ estimation differs from the value obtained for a test particle.}
    \label{tab:rc_mp_eq_30mj}
\end{table*}

    \begin{table*}
    \centering
    \begin{tabular}{| c c c c c |}
    \multicolumn{5}{c}{Critical semi-major axis ($m_p = 0, 1, 30$ M$_\textrm{J}$)} \\
    \hline 
    \multicolumn{3}{|c}{} & \multicolumn{1}{c}{$\mu$} & \multicolumn{1}{c|}{} \\
            &    & 0.20    & 0.50   & 0.90 \\
    $e$ & $m$  (M$_\textrm{J}$) & \multicolumn{2}{c}{} & \multicolumn{1}{c|}{} \\   
    \cline{3-5}      
    0.0     & 30 & $0.368 \pm 0.002$  & -                  & - \\
            &  1 & $0.378 \pm 0.002$  & -                  & -   \\
            &  0 & $0.380 \pm 0.002$  & -                  & - \\
    \cline{3-5}
    0.2     & 30 & -                & $0.208 \pm 0.002$    & -  \\
            &  1 & -                & $0.208 \pm 0.002$    & -  \\
            &  0 & -                & $0.210 \pm 0.002$    & -  \\
    \cline{3-5}
    0.6     & 30 & $0.114 \pm 0.002$   & -                 & $0.048 \pm 0.002$  \\
            &  1 & $0.118 \pm 0.002$   & -                 & $0.048 \pm 0.002$    \\
            &  0 & $0.118 \pm 0.002$   & -                 & $0.049 \pm 0.002$ \\
    \hline
    \end{tabular}
    \caption{{\bf High precision estimates of the critical semi-major axis $a_c$ for the $(e, \mu)$ values reported in the highlighted entries of table \ref{tab:rc_mp_eq_30mj}
    .}}
    \label{tab:rc_mp_eq_30mj_zoom}
\end{table*}

    For a given binary, the planetary motion depends mainly on the initial semi-major axis $a$ of the planetary orbit. In general, a planet with a small a, i.e. revolving close to the primary star, has a stable orbit. We apply here the idea, introduced by H99, of critical semi-major axis $a_c$ as threshold from stability to instability, to evaluate how orbital stability depends on the mass ratio $\mu$ and the eccentricity $e$ of the binary star system. We notice that $a_c$ is well defined only if the boundary between the region of stability (where $a \le a_c$) and instability where ($a > a_c$) is sharp. In Tables  \ref{tab:stability_p} and \ref{tab:stability_a} we report the results for a binary star system with $e = 0.5$  and $\mu = 0.5$; in Table \ref{tab:stability_p} the binary is initially at periapse, in Table \ref{tab:stability_a} the binary is initially at apoapse. Our choice of considering eight values ($0$, $\pi/4$, $\pi/2$, $(3/4) \pi$, $\pi$, $(5/4) \pi$, $(3/2) \pi$, $(7/4)\pi$, $2\pi$) for the initial longitude of the planet allows a robust estimate of $a_c = 0.12$, that is the same value obtained by H99. We characterize the stable orbits by the value of $\Delta r$ (see end of Sect. \ref{sec:methods}). For stable orbits, in some cases we estimate a value of $\Delta r$ greater than $5\%$, which define a marginally stable orbits, according to the classification introduced for the first time by \cite{Musielak2005}. For unstable orbits, we evaluate whether i) there is a collision with the primary star (CA) or ii) with the secondary star (CB), or iii) if the planet is ejected away from the binary star system (EJ). In the particular case of a binary with $\mu = 0.5$ and $e = 0.5$ (Tables \ref{tab:stability_p} and \ref{tab:stability_a}), most of the instabilities correspond to collisions with the primary star, but there are also ejections and a few collisions with the secondary star. We obtain different survival times than H99, likely due to the different, and much more accurate, integration scheme we adopted. We find some stable orbits beyond the critical semi-major axis ($a_c = 0.12$): one is quoted in Table \ref{tab:stability_p} with $a = 0.13$, four are quoted in Table \ref{tab:stability_a} with $a = 0.13, 0.14$. We do not find mean motion resonances in these orbits.
    
    The region of stability ($a < a_c$), in which all the orbits are stable at any longitude, is well defined for all the binaries systems that we considered. The dependence of the critical semi-major axis $a_c$ on the mass ratio $\mu$ (from $0.1$ to $0.9$) and eccentricity $e$ (from $0$ to $0.8$) of the binary star system for a planet of negligible mass and equal of $1$ and $30$ Jupiter mass is reported in Table \ref{tab:rc_mp_eq_30mj}. The results for planets of negligible mass are in agreement with H99. In most cases, we find that the value of $a_c$ does not depend on the mass of the planet, with the exception of some entries, that show different values of $a_c$, especially for $30 M_J$. However this result is not reported in Table \ref{tab:rc_mp_eq_30mj} with a high statistical significance. The large error is due to the fact that $a_c$ values are using a $\Delta a_c$ step equal to $0.01$ (see Table \ref{tab:stability_p} and \ref{tab:stability_a}). In order to obtain a greater resolution, we recalculated $a_c$ in four cases  using a smaller $\Delta a_c$ step (Table \ref{tab:rc_mp_eq_30mj_zoom}). The new outcomes confirm a significant dependence for planets of mass equal to $30 M_j$. We find that a mass of the planet equal to $\ M_j$ does not produce significant effects in our simulations, with the exception of the $(e, \mu) = (0.2, 0.50)$ configuration, where however the significance is quite low. In all the $81$ binary configurations we find that a fraction of the orbits are stable beyond the critical semi-major axis. Considering a narrow neighborhood of $a_c$, the number of stable orbits strongly depends on the $(\mu, e)$ value. In the case reported in Table \ref{tab:stability_p} and \ref{tab:stability_a} ($\mu = 0.5$, $e = 0.5$, $a_c = 0.12$) there are $4$ stable orbits with $a = 0.13$; the fraction of stable orbits in this case is $4$ over $16$ ($25 \%$). The mean number of stable orbits beyond $a_c$ over all binary configurations does not depend significantly on the mass of the planet. Averaging over all binary configurations, the fraction of stable orbits in a narrow neighborhood of $a_c$ is $55 \%$.

\section{Discussion}
\label{sec:discussion}

    In our numerical study, we simulated more than ten thousand S-type orbits of planets in a binary star system, using a high precision, $15$th order, integration scheme. We estimated the \textit{critical} semi-major axis, which defines the size of the stable regions, as a function of the mass ratio and orbital eccentricity of the binary star system hosting the planet, finding results in agreement with H99 for planets with negligible mass. 
    
    We also studied the motion for planets with a mass of $1$ and $30$ Jupiter. This is a reasonable choice, because it covers the observational data. Indeed, looking at the distribution of the masses measured in binary systems (Fig. \ref{fig:masses}), we note that the majority of the planets have a mass equal to about two hundredths of the Jupiter mass, with a distribution of masses extending up to 16.1 Jupiter mass \citep{Liu2008}. Considering all the planetary systems (i.e. including those around single stars), the largest mass reported in the data archive (exoplanets.org) is equal to 22.6 Jupiter mass \citep{Sato2010}; in this extreme case the small mass tertiary around the binary falls in the brown-dwarf regime.
    
    \cite{Marzari2016}, using the FMA method to select stable orbits, found a significant numbers stable planetary orbits in binaries beyond this critical distance from the primary star. In general, this result does not disagree with our simulations, since we find in many cases stable orbits beyond the critical distance, as shown for example in Tables \ref{tab:stability_p} and \ref{tab:stability_a} for two stars of equal mass and eccentricity equal $0.5$. For very massive objects ($30 Mj$) we find differences in the  $a_c$ estimation of the order of $3 \%$ compared with test (negligible mass) particles (see Table \ref{tab:rc_mp_eq_30mj_zoom}). It is worth noting that we find these dependence on the planetary mass for low binary mass ratio $\mu$. This is expected, since in these cases the mass of the secondary star is smaller than that of the primary star, so where the mass of the planet plays a more important role. 
    
    For all the stable orbits  we estimated the spread ($\Delta r$) of the distance of the planet from the primary star with respect to its initial value, showing that a fraction of these orbits is only \textit{marginally} stable, according with the classification of \cite{Musielak2005}. Our data analysis also produces statistics of planet-star collisions and planetary ejections away from the binary, in the cases of planet mass equal to $0,1, 30$ Jupiter masses. In the particular case of stars of equal mass and moving on an $e=0.5$ orbit, for all planet masses studied here, the largest fraction of unstable orbits result into a collision of the planet onto the primary star, followed by ejection and collisions onto the secondary. A planet-star collision might result in the pollution of stars with planetary material, which, in principle, would be detectable by accurate spectroscopic measurements. The measurements of iron abundance for a sample of $23$ wide binaries reported by \cite{Desidera2004} seem to rule out a contamination with a mass of iron greater than $1$ Earth Mass during the Main Sequence lifetime of the stars. The instability of planetary orbits in double star systems with a final fate as  ejection of the planet away from the binary system has an relevance in increasing the number of freely-floating planets as detected by gravitational microlensing \citep{Sumi2011}.
    
    In terms of the parameter space, we made some assumptions. We considered binary systems hosting a single planet moving on an unperturbed circular co-planar orbit around the primary star.  For the data analysis, we implemented a code based on standard Python tools (\emph{numpy}, \emph{pandas}). A joint application of the high-order integration code and our data analysis tools on a High Performance Computing (HPC) platform will allow a significant extension of the space of parameters to investigate. An interesting approach to analyse the large amount of output data that will be produced is based on Machine Learning techniques (see e.g. \cite{Tamayo2016}).
    
    An interesting perspective is to investigate our results in the scenario of the subresonances theory. Indeed, the planet experiences resonant perturbations from the companion star. Starting from the observation that planet may be dislodged from its host star if it is simultaneously affected by two or more resonances, \cite{Mudryk2006} found that overlap between subresonances lying within mean-motion resonances can explain the boundary of orbital stability within binary systems.  
    
\begin{acknowledgments}
   The authors are grateful to the anonymous referee for the useful discussion, and to the Associate Editor of Astrophysics and Space Science Prof. Krzysztof Gozdziewski for the useful suggestions.
\end{acknowledgments}

\section*{Data availability}
Some of the data underlying this article are available in the github repository  \url{https://github.com/giovixo/binary_archive}, and can be accessed with the DOI \url{https://zenodo.org/record/4014613}.

\bibliographystyle{spr-mp-nameyear-cnd}
\bibliography{decesare_ref} 

\end{document}